**Evaluating Incidence and Impact Estimates of the Coronavirus Outbreak from Wuhan before Lockdown**

Mai He, M.D., Ph.D.,[1] Li Li, M.A.[2], Louis P. Dehner, M.D.[1]

1. Department of Pathology and Immunology, Washington University School of Medicine in St. Louis, St. Louis, MO 63110, USA
2. AT&T, Bedminster, NJ 07921, USA

**Corresponding author:**

Mai He, M.D., Ph.D.

Department of Pathology and Immunology,

Washington University School of Medicine in St. Louis,

St. Louis, MO 63110, USA

Phone: (314) 273-1328

Email: Maihe@wustl.edu

**Abstract**

Background:

Wuhan, China was the original epicenter of COVID-19 pandemic. The goal of the current study is to understand the infection transmission dynamics before intervention measures were taken, such as issuing a lockdown for the city and other social distancing policies.

Methods:

Data and key events were searched through pubmed for medical literature and internet for Chinese government announcements and Chinese media reports. Epidemiological data including R0 and infection were calculated using data extracted from variety of data sources.

Results:

We established a timeline emphasizing evidence of human-to-human transmission. By January 1, 2020, Chinese authorities had been presented convincing evidence of human-to-human transmission; however, it was until January 20, 2020 that this information was shared with the public. Our study estimated that there would have been 10,989 total infected cases if interventions were taken on January 2, 2020, vs 239,875 cases when lockdown was put in place on January 23, 2020.

Conclusions:

China's withholding of key information about the 2020 COVID-19 pandemic and its delayed response ultimately led to the largest public health crisis of this century and could have been avoided with earlier countermeasures.

**Introduction**

The COVID-19 outbreak that originated from Wuhan, a major transportation hub of China, had caused more than eight million confirmed cases and 444,563 deaths in the world as of June 17, 2020 [1]. Multiple countries reported having their first cases imported from Wuhan or by a returning visitor to Wuhan [2-7]. Thus, it is important to understand the infection transmission dynamics within the epicenter where the virus first originated, especially before intervention measures were taken such as issuing a lockdown of the city and other social distancing policies. In spite of the worldwide interest in understanding the initial phase of the pandemic in Wuhan, researchers' efforts have been hampered by lack of transparency including the lack of comprehensive and reliable data from this period.

By the time of the lockdown on January 23,2020, the Chinese official report for the COVID infection in Wuhan is 495 cases. However, media reports from China suggest that the official confirmed cases in Wuhan before and around the lockdown were significantly underreported. This under-reporting seems to be due to multiple factors, such as limitation of testing and surveillance capability [8,9]; limited quantity of test kits provided to local hospitals [10] and generally low testing accuracy (low sensitivity, 30-60%) [11]. More importantly, government policies and intervention, such as the different versions of diagnostic criteria, seem to have a heavy influence on the information sharing and preparation for the pandemic [12].

The current study investigates the early outbreak through search and screening into various types information, including interviews with those who were in direct contact with patients, Chinese media reports and government reports. We found that the key information of human to human transmission had not been reported by the Chinese government for at least three weeks, despite convincing evidence. We provide our estimates of the impact of this delay of three weeks and other epidemiological parameters related to COVID-19 outbreak in Wuhan, before the lockdown and other public quarantine measures on January 23, 2020.

**Methods**

*Data Collection*

In an effort to reconstruct what happened before and during the lockdown in Wuhan, we screened media reports published in mainland China from January to April 2020, covering the details about various events and measures taken by local hospitals and government on various levels concerning the Wuhan lockdown. We have complied the epidemic relevant events into a timeline so that the scientific community will have information that can potentially impact understanding the dynamics of the COVID-19 virus in Wuhan, China.

We searched and profiled data through the following different information channels.

1). Pubmed search using keyword "COVID-19". Search results were screened for daily incidence by PCR diagnosis from admitted patients. Two references were found [13, 14]. [Appendix A, column B].

2). Screening of published Chinese media reports with recounts from local physicians and hospitals. Search results were screened for daily and total hospitalization cases including the PCR confirmed cases and patients with high suspicion of COVID infection with severe symptoms. We found the following Chinese media reports and government announcements [15-19] [Appendix A, column C].

3). Daily new incidences with a typical pattern of COVID infection on CT images with features strongly correlated to the known COVID pattern in two radiology departments in two hospitals in Wuhan from media reports [9] [Appendix A, column H].

*Data Classifications*

We have created three data categories for our calculations:

A: Hospitalized patients with COVID infection confirmed by PCR in Wuhan citywide.

B: Hospitalized patients due to COVID infection in Wuhan citywide (including highly suspicious clinical cases not yet confirmed by PCR or local diagnosis criteria).

C: Daily symptomatic incidences based on daily cases with COVID patterns identified by radiologists from CT reading rooms from two similarly sized Grade III Level A (tertiary) hospitals.

**Results**

*I.    Timeline of COVID-19 Outbreak Before Lockdown and Chinese Government's Reaction*

A brief timeline of COVID-19 outbreak, focusing on features of human-to-human transmission, before Wuhan lockdown on January 23, 2020, is listed in Figure 1 [20, 21] (for a more comprehensive timeline and its references, see Appendix B).

*I.I    Patients in family clusters suggestive of human to human transmission*

On December 27, 2019, an elderly couple presenting with fever and dyspnea was admitted to Wuhan Xinhua Hospital ICU. On the next day, the son of this couple was called to Xinhua Hospital. The CT imaging of all three in the family showed similar findings. This family cluster demonstrated an obvious clue of "human to human transmission." ICU chief, Dr. Zhang Jixian reported these three cases to the hospital leadership. On December 29, the vice president of the Xinhua Hospital, Dr. Xia Wenguang called for a clinical conference on seven similar cases, including this family cluster of three. On the same day, Dr. Xia reported to the City and Provincial Health Commissions and CDCs. He asked for starting emergency protocol activation.

On December 29, 2019, the ER in Wuhan Central Hospital reported four similar cases to the Hospital Public Health Office, which reported to local county CDC. Among these patients, there was a pair of mother-son, another example of human-to-human transmission.

On 12/30/2019, one physician in Xinhua Hospital reported symptoms. This was confirmed on Jan 11, and she is the 2nd case of a medical staff infected in Xinhua Hospital.

### I.II. *Next-generation sequencing (NGS) of airway samples suggestive of SARS-like coronavirus*

On December 27.2019, the NGS result of patient A was orally reported back to Central Hospital by Vision Medicals (Guangzhou, China) by phone, it was claimed to be a coronavirus that is "Bat SARS like". The sequencing results of patient A were forwarded to Wuhan Institute of Virology (WIV), and WIV confirmed the high similarity to SARS-like coronavirus from bat.

On December 30, 2019, Beijing CapitalBio MedLab (北京博奥医学检验所) reported sequencing results of patient Chen 2 as "highly suspicious for SARS" to the Central Hospital. Dr. Ai Fen, ER Chief in Central Hospital, reported sequencing results to Hospital Public Health Office and Hospital Infectious Disease Department. She also forwarded the report to ER doctor's Wechat, which triggered a chain of events spreading the warning, including message sent out by Dr. Li Wenliang. On January 1, 2020, BGI group (Huada Gene) reported sequencing results of 3 cases of pneumonia in Wuhan which harbored viral genome sequences showing 80% similarity to SARS virus to Wuhan City Health Commission.

*I.III. Non-acknowledgement of human-to-human transmission and intervention in case reporting*

As Table 1 demonstrates, by December 30, 2019, there had been convincing evidence of human-to-human transmission, including clinical presentation of patient clusters and viral sequencing results. Despite this, on December 30,2019, the Wuhan City Health Commission issued two emergency policies, the 2nd one stating that "no one or institution is allowed to disclose information about the pneumonia (of unknown etiology) without authorization." On December 31, 2019, the Wuhan City Health Commission announced that recent pneumonia cases seen in some hospitals were related to the Wet Market; no obvious findings of "human to human transmission" and no infection of medical staff.

On December 31, 2019, the Chinese National Health Commission dispatched a panel of experts to Wuhan for investigation. After visiting Wuhan, they announced no evidence of "human to human transmission" on January 10, 2020. It was not until January 20, 2020 that the National Health Commission announced the human to human transmission by COVID-19. Three weeks passed from the emerging evidence of human-to-human transmission.

During this period, Chinese physicians sensed the risk of outbreaks and reported promptly; Chinese researchers also cultured and isolated the virus and obtained the viral genome sequence. On the other hand, while the Chinese government started to communicate with international organizations, the severity was under-reported and "human to human transmission" was denied. Other counterproductive measures taken by the Chinese government included, but were not limited to, prohibiting medical staff from releasing information to public, prohibiting discussions among medical staff on the "pneumonia of unknown etiology", prohibiting staff from wearing PPE (personal protection equipment), prohibiting 3rd party labs from doing viral sequencing, making diagnostic criteria difficult for diagnosis confirmation, deleting reported cases, disciplinary action against "whistle blowers" and closure of the lab which shared viral genome information with the international community.

Wuhan local and the Chinese National CDC has implemented at least seven different versions of diagnosis criteria since the outbreak, thus making it hard to identify and understand the infection number in a consistent manner [12]. If taking infection with mild or no symptoms into account, a study estimated that 86% of infected individuals were not documented in China before January 23, 2020 [22]. Policies and interventions for reporting (holding cases for 12 days) made the confirmed cases artificially low, as reported by mainland Chinese media [15,16, 23].

### I.IV.   *Reaction of Taiwan and Hong Kong governments*

By comparison, the Hong Kong and Taiwan governments were closely monitoring and preparing for the coming epidemic during this critical time. On December 31, 2019, Taiwan inquired of China and the WHO for details about the "Atypical Pneumonia patients quarantined" with a strong suspicion of human to human transmission. Taiwan started border infections screening on direct flights from Wuhan to Taiwan. On December 31, 2019, Hong Kong media reported the presence of "new SARS" in Wuhan. The Taiwan and Hong Kong governments had press conference almost daily on the epidemic.

## II. Estimation of Epidemic in Wuhan

### II.I. Calculation of $R_0$

Assuming the ratio of symptomatic cases in the total infected population is constant, and the proportion of symptomatic patients seeking medical help remains constant over the time, we use the growth rate of hospitalized people, PCR test positives or CT positives in our dataset as the proxy for the growth rate of the infected population.

$R_0$ represents the average number of secondary cases that result from the introduction of a single infectious case in a totally susceptible population during the infectious period. Wallinga and Lipsitch have pointed out that the exponential growth rate (r) during the early phase of the outbreak can be used to calculate the $R_0$ by using the formula $R_0 = 1/M(-r)$, where M is the moment generating functions of the serial interval (SI) distribution of the disease [24].

We assumed a gamma distribution for the serial interval, with 7.5 as the mean and 3.4 as standard deviation, based on a prior study of Wuhan data of the initial epidemic stage [13]. R package "$R_0$" is used to provide estimates of the $R_0$ and its 95% confidence interval [25].

In the Feb. 3 speech, Chinese President Xi told China's most powerful leaders that he had "continuously given verbal and written instructions" since Jan 7, and he had personally ordered the quarantine of about 60 million people in Hubei province later in January. We found PCR reported cases decreased after January 9, 2020, while no countermeasure had been taken by then. The decrease in reporting could well be driven by political reasons. Thus, we utilize (dataset A) PCR results from December 1, 2019 to January 8, 2020 (the growing portion which is less subject to political intervention) [Figure 2], $R_0$ is estimated to be 3.12, with 95% confidence interval as (2.69, 3.64) [Table 2].

Patients without symptoms are highly unlikely to be tested for COVID in Wuhan, they will get CT scans only if they have symptoms that are either severe or causing respiratory distress; thus, the CT rooms readings provide samples with severe symptoms. The CT data (data set C) fit well with a Poisson log linear regression model for the daily new incidence. The coefficient of the Poisson model is 0.179 (0.151, 0.207), which implies 3.565 doubling time and higher $R_0$ than the PCR based $R_0$. However, due to the limited size of this dataset, this study will mainly use the PCR based $R_0$ for estimation.

Based on field observation, around 60,000 more people visited febrile clinics in Wuhan from January 22 to January 27, 2020 with symptoms, and 3,883 were observed at clinics who could not be admitted due to lack of beds [Appendix A, column I]. Chinese media reported that shortage of hospital beds wasn't eased until February 15, 2020. The growth of the hospitalized population was subject to the capability of Wuhan for adding beds dedicated to COVID patients. Thus, we are not going to use the daily incremental hospitalization for the $R_0$ estimation, since it could likely produce an underestimate.

## II.II. *Estimate of incidence in Wuhan based on the SEIR model*

Susceptible–exposed–infectious–removed (SEIR) model is applied to estimate the incidences. The following epidemiologic parameters are assumed in the estimation [13].

1. December 1, 2019 as the initial outbreak time, January 23, 2020 is the 54$^{th}$ day.
2. 5.2 days as the mean incubation duration
3. 2.3 days as the mean infection duration
4. 11.5 days as the median time from initial hospital admittance to discharge [26]
5. 2% case fatality rate (Chinese official figure of the early stage in Wuhan)

6. 14 million population in Wuhan before lockdown [27]

With the estimated 3.12 $R_0$ based on the PCR results and the SEIR model, it is estimated that by January 23, 2020, Wuhan had 239,875 infected people, among them 149,170 were in the stage of incubation and were infectious. The new daily latent and infectious cases are estimated to be 23,774, 7,148 respectively on January 23, 2020. [Figures 3A]

*II.III. Hospital admission rate: government official report vs estimation*

The Chinese official report for Wuhan suggested a 100% admission rate for COVID patients diagnosed in Wuhan, China [28]. However, it was very hard to secure a bed around the lockdown period in Wuhan, and many residents were told to go back home and self-quarantine [29]. By December 31, 2019, the officially reported admitted number is 46, our analysis estimates a total of 333 (105, 1,219) infectious cases by then, which implies 13% admission rate (46/333). On January 20, 2020, it was reported the 2,000 beds dedicated to COVID were all occupied, we estimated 14,804 (2,485,104,298) infectious cases cumulated, and a 13.5% admission rate. The above hospital admission ratio was based on the estimates of infectious cases, many infected but no longer infectious cases needed medical care as well.

*III.  Estimation of Impacts of the Delay of the Three Weeks*

The period of the three weeks before the lockdown was a crucial time in the early stage of the epidemic. If countermeasures would have been taken on January 2, 2020, immediately after the authorities were informed about the human-to-human transmission, the epidemic development would have taken a completely different trajectory. If we assume the intervention reduces the daily contacts of the infectious by 2/3, our study estimated that there would have been 10,989 total infected cases, vs 239,875 cases when lockdown was taken place on Jan 23, 2020. The group of people with infectivity (infectious+ latent) would be reduced from 149,170 to 1,916, and the daily new cases of the infectible group would be reduced from 30,922 to 10 on January 23, 2020 [Table 3] [Figure 3B]. This demonstrates that if countermeasures would have been taken three weeks earlier, the epidemic would have been largely contained by January 23, 2020.

By the time the authorities locked down Wuhan, about five million people (about 5/14 of the total Wuhan population) had left Wuhan, resulting in a potential outflow of 41,008 exposed cases and 12,267 infectious cases, contributing directly to one of the most serious public health crises of this century.

The above estimations assume the intervention reduces the daily contacts of the infectious by 2/3. After the long delay, Wuhan took a drastic lockdown measure to contain the virus spread, which reduced the daily contacts of the infectious by more than 2/3. If a strict intervention would have been applied on January 2, 2020, we would have seen an even larger decline in the epidemic spread.

**Discussion**

*Potential impact of the delay of the critical three weeks*

Some other scholars have done hypothetical studies of countermeasures. As Yang et al. suggested, using a modified SEIR and AI prediction of the epidemic trend of COVID-19 in China, "If the introduction of interventions was delayed by five days, the transmission coefficient would have been much greater due to the increase in the average number of contacts with an infected person daily." They also pointed out that a five-day delay in implementation would have increased the epidemic size in mainland China three-fold [30]. Lai S and associates pointed out that if interventions in China could have been conducted one week, two weeks, or three weeks earlier, cases could have been dramatically reduced by 66% (IQR 50% - 82%), 86% (81% - 90%), or 95% (93% - 97%), respectively [31].

Our study has demonstrated that the delay was real, there is 21 day interval between December 31 and January 20 when China acknowledged the human to human transmission, and 24 days to January 23 when lockdown took place. The experience of Hong Kong and Taiwan has shown that the infection can be contained if actions are taken promptly [32,33].

Based on our media report screening, we did not identify any public quarantine measure taken by the government in Wuhan before January 23, 2020. Instead, communities were encouraged to continue hosting large scale gatherings to celebrate the lunar New Year. On January 18, 2020, over 40 thousand families attended the Ten Thousand Families Banquet at the Baibuting Community, none of the attendees wore masks. Three days before this event, some organizers asked to cancel the banquet, but the request was not approved [34]. In the meantime, the authorities took steps to underreport the incidence and infectivity of COVID-19 to China and the world.

*Emigration and estimation*

In this study, we did not incorporate the dynamics of the population in Wuhan. About five million people left Wuhan before the lockdown, with a large portion happening during the week before Chinese New Year's Day (January 25, 2020). The infected among this emigrant group got in contact with population outside Wuhan, making the susceptible number larger. Also, the higher frequency of interpersonal contacts of these emigrants before the Chinese New

Year's Day will result in a higher $R_0$. Hence, the estimation of the infected presented in this paper is likely to be an underestimate.

### *The start date of the outbreak*

In this study, we assumed the outbreak first took place on December 1, 2019. Some studies suggest an earlier date. However, Ma J reported that according to China's internal government's data, the first case of COVID-19 confirmed case could be traced back to November 17, 2019 [35]. The Epidemiology Working Group for NCIP Epidemic Response and Chinese Center for Disease Control and Prevention, retrospectively reported 104 cases by December 31, 2019 based on data extracted from China's Infectious Disease Information System [36]. On September 18, 2019, Chinese media reported a drill by Wuhan Customs, including a case scenario of "novel coronavirus infection" [37]. The above information suggests possibilities that are not known to public.

### *$R_0$ estimation and modeling*

The epidemiological parameters are important for estimating the potential impact of the at least three-weeks delay in response by Chinese government. $R_0$ for Wuhan has been studied by other groups. Li Q et al. from the Chinese CDC and other institutions studied the Wuhan epidemic curve up to January 4, 2020, with estimation of R0 as 2.2 [13]. Zhao S et al. applied an exponential growth model, adjusted report rate of the Chinese official data, and estimated the R0 to be between 2.24 to 3.58 [38]. A study by Sanche S and associates looked into Wuhan travelers who were confirmed infected in other provinces and estimated the Wuhan outbreak had a R0 of 5.7 (3.8–8.9) [26]. Hou J et al estimated the R0 in Wuhan to be 4.90 during the local epidemic period from December 1, 2019 to January 9, 2020, and 3.00 during the long-distance spread period due to the Spring Festival travel rush from January 10 to 22 [39]. Read JM et a.l estimated that only 5.1% of infections in Wuhan's early outbreak were identified, and R0 as 3.11 between January 1 to January 21 [40]. Wu JT et al studied cases exported from Wuhan to other cities outside of Mainland China using modelling and estimated R0 to be 2.68 [41].

Despite our efforts to extract clinical operation-related data from various types of reports, the data we have collected are still fragmented and largely incomplete due to the lack of transparency in China. Nevertheless, our PCR based $R_0$ estimate 3.12 with 95% C.I (2.69, 3.64) is compatible with the previous studies. As shown in table 5, if the actual $R_0$ is higher than 3.12, the case reduction brought by intervention will be more significant; if the actual $R_0$ is lower, the epidemic would have been contained by January 23, 2020.

Wuhan is not a homogenous city, the SEIR model at most can be used to give a crude oversimplified mathematical modelling of real-life events and estimates of the magnitude of infection. Given the nature of MCMC methods, the samples/incidences generated in our case best serve as a rough estimate of the size of the Wuhan epidemic and indicator of discrepancy from official figures. We thus acknowledge the limitations to the accuracy of our measures.

**Conclusion**

We investigate the COVID-19 outbreak related clinical operations in Wuhan by drawing a timeline of the COVID-19 outbreak before lockdown, focusing on "human to human transmission." This timeline demonstrates that the Chinese government did not provide critical information it possessed of human to human transmission for at least three weeks, while intervening in the case reporting. Without the three-week delay, the largest public health crisis of this century could have been avoided or controlled to a much greater degree.

**Acknowledgements:**

To Chinese physicians and all health care professionals, among them are my classmates, and to media reporters, for their courage and professionalism.

We thank Professor Lucia Dunn at the Ohio State University for her help on data analysis and language editing of this manuscript.

**Funding:**

None

**Declaration of interest:**

The authors declare no conflict of interest.

**Figures legends:**

Figure 1. Brief timeline of COVID-19 outbreak in Wuhan, China, before lockdown on January 23, 2020, emphasizing events related "human to human transmission" .

Figure 2. Calculation of R0 based on dataset A up to January 8, 2020.

Figure 3A: Estimation of COVID-19 Infection with Intervention Taken on Jan 23, 2020.

Figure 3B: Estimation of Infection with Intervention Taken on Jan 2, 2020.

Table 1. Evidence of "human to human" transmission in Wuhan, China, by December-31, 2019 [20,21]

| Hospital | Case | Dates |
|---|---|---|
| Xinhua Hospital | The case of family of 3 | Dec 27 and 28, 2019 |
| Central Hospital | Sequencing result of "Bat SARS like" novel coronavirus reported by Guang Zhou (Vision Medical) from patient A (Dec-18-2019) | Admitted on Dec 18, 2019; Sampled on Dec 24, 2019; Sequencing result reported on December 30, 2019 |
| Central Hospital | The case of mother-son pair | December 29, 2019 |
| Central Hospital | Beijing (CapitalBio MedLab) reported sequencing results of patient Chen 2 as "highly suspicious for SARS". | Admitted on December 27, 2019; Reported on December 30, 2019 |
| Xinhua Hospital | One physician in Xinhua Hospital reported symptoms. This was confirmed on Jan 11, 2020 and she is the 2nd case of medical staff infection in Xinhua Hospital. | December 30, 2019 |

Table 2: $R_0$ estimation

| | |
|---|---|
| Dataset A: PCR Incidence (December 1, 2019 - January 21, 2020) | 1.62 (1.53, 1.72) |
| Dataset A: PCR Incidence (December 1, 2019 – January 8, 2020) | 3.12 (2.69, 3.64) |
| Dataset B: Hospitalized (December 1, 2019 – January 21, 2020) | 2.28 ( 2.16, 2.41) |
| Dataset C: CT (January 1, 2020 – January 21, 2020) | 3.70 (3.36, 4.06) |

| Table 3: Infection of COVID-19 from Wuhan on January 23, 2020 ||||||||||
| R0 | Intervention | Latent (incubation) | Infectious | Infected but no longer infectious | Recovered | Died | Total infected | Total infected staying Wuhan | Total infected out of Wuhan |
| --- | --- | --- | --- | --- | --- | --- | --- | --- | --- |
| 3.12 (2.69, 3.64) | January 23, 2020 | 114,823 (15,935, 900,700) | 34,347 (5,021, 272,953) | 71,193 (12,423, 501,280) | 19,101 (3,796, 129,551) | 411 (81, 2,608) | 239,875 (37,256, 1,807,092) | 154,205 (23,950, 1,161,702) | 85,670 (13,306, 645,390) |
| 3.12 (2.69, 3.64) | January 2, 2020 | 1,333 (258, 8,421) | 583 (118, 3,508) | 5,807 (1,536, 26,879) | 3,199 (910, 13,639) | 67 (19, 286) | 10,989 (2,841, 52,724) | 7,064 (1,826, 33,894) | 3,925 (1,015, 18,830) |

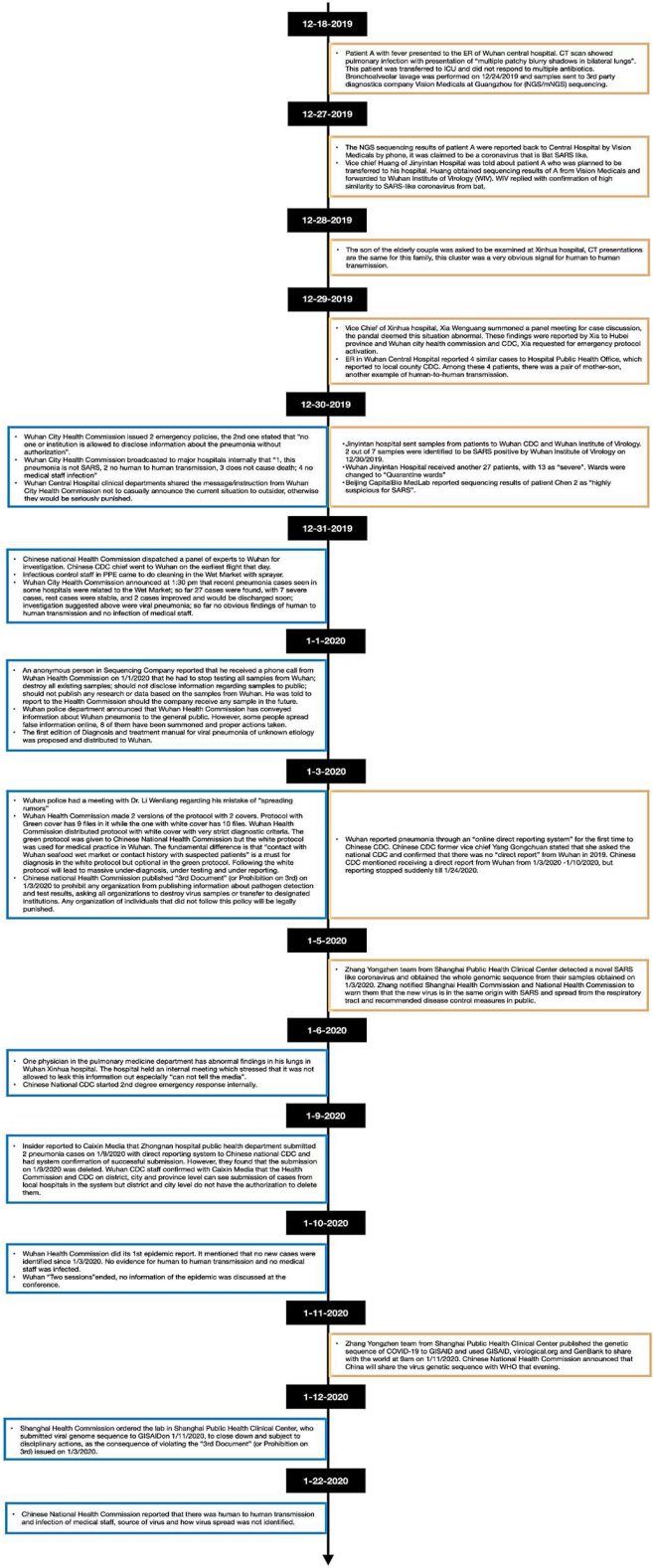

**Figure 1. Brief timeline of COVID-19 outbreak in Wuhan, China, before lockdown on January 23, 2020, emphasizing events related "human to human transmission."**

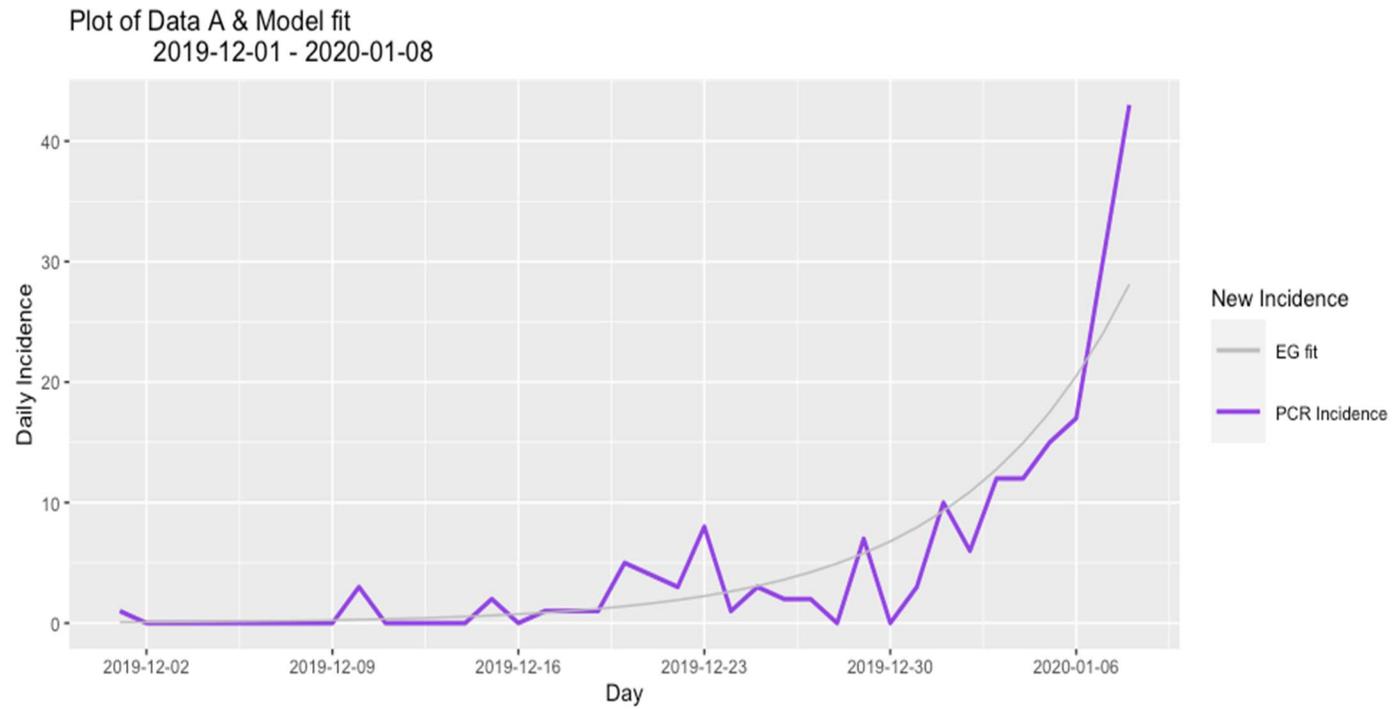

**Figure 2. Calculation of R0 based on dataset A up to January 8, 2020.**

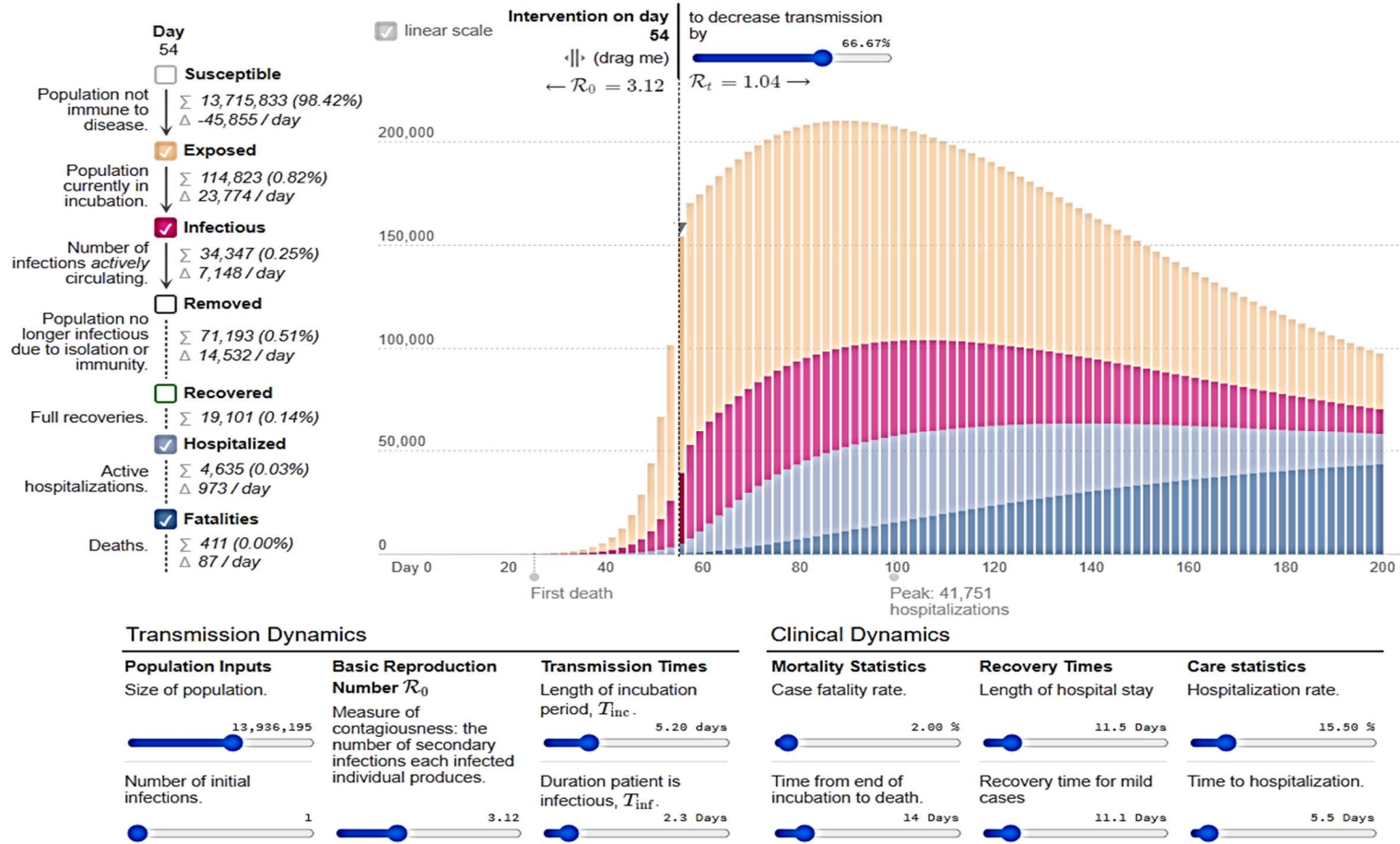

Note: calculation was done with the epidemic calculator. http://gabgoh.github.io/COVID/index.html

**Figure 3A: Estimation of COVID-19 Infection with Intervention Taken on Jan 23, 2020.**

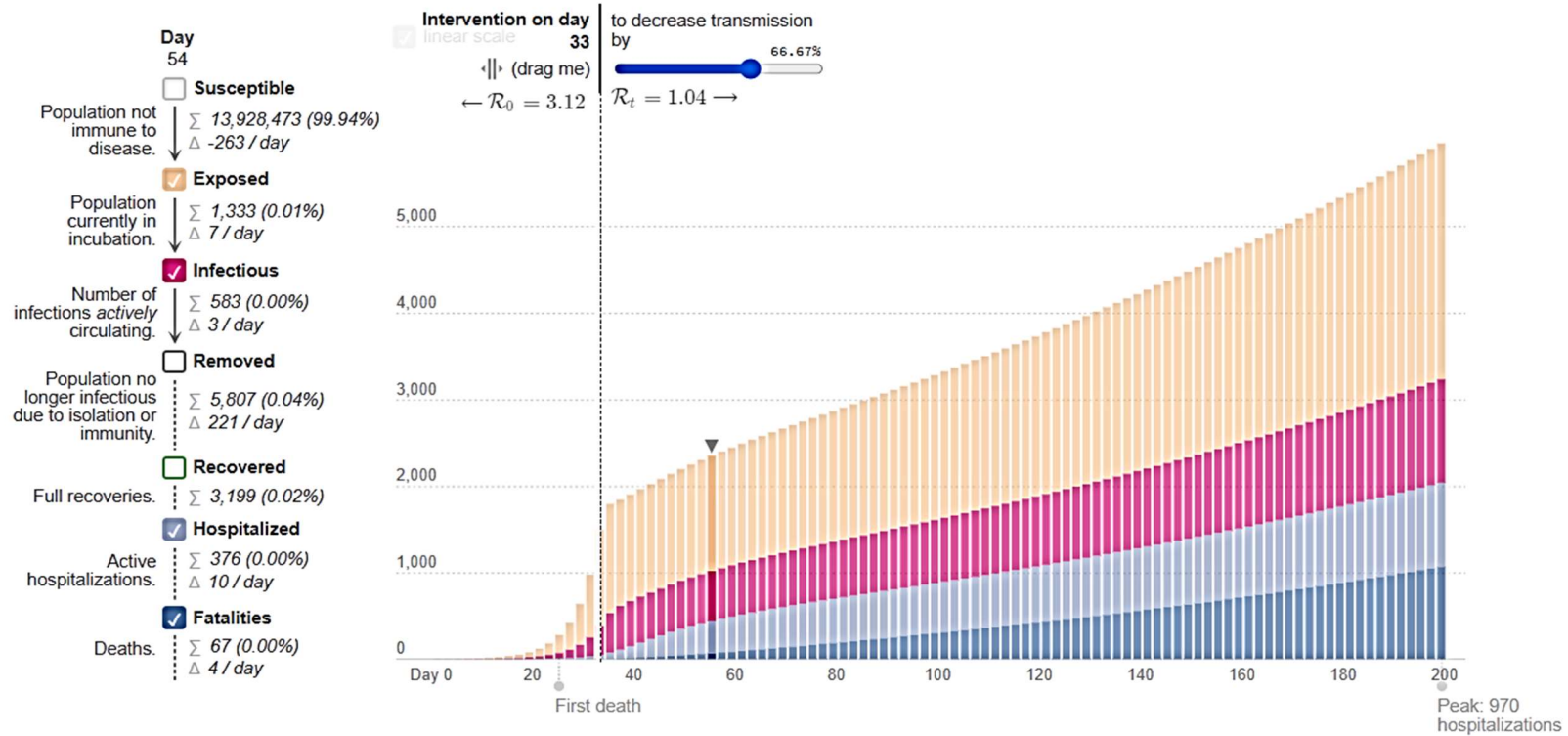

Note: calculation was done with the epidemic calculator. http://gabgoh.github.io/COVID/index.html

**Figure 3B: Estimation of Infection with Intervention Taken on Jan 2, 2020.**

# Supplementary materials

# Table of Contents

# Appendix A. Data

| A | B | C | D | E | F | G | H | I |
|---|---|---|---|---|---|---|---|---|
| Date | New Confirmed Incidence by PCR [13,14] | Hospitalized (new incidence) Inferred to be positive cases [15-19] | Hospitalization (new incidence) [including confirmed and inferred] | Symptomatic cases at Clinic (inferred to be positive for COVID) | Total Confirmed by PCR (total of Column B by date) **Dataset A** | Total Hospitalized (total of column D by date) **Dataset B** | CT Clinical diagnosis daily Incidence **Dataset C** | Total Symptomatic cases at Clinic (inferred to be positive for COVID) |
| 12/1/2019 | 1 | | 1 | | 1 | 1 | | |
| 12/10/2019 | 3 | | 3 | | 4 | 4 | | |
| 12/15/2019 | 2 | | 2 | | 6 | 6 | | |
| 12/17/2019 | 1 | | 1 | | 7 | 7 | | |
| 12/18/2019 | 1 | | 1 | | 8 | 8 | | |
| 12/19/2019 | 1 | | 1 | | 9 | 9 | | |
| 12/20/2019 | 5 | | 5 | | 14 | 14 | | |
| 12/21/2019 | 4 | | 4 | | 18 | 18 | | |
| 12/22/2019 | 3 | | 3 | | 21 | 21 | | |
| 12/23/2019 | 8 | | 8 | | 29 | 29 | | |
| 12/24/2019 | 1 | | 1 | | 30 | 30 | | |
| 12/25/2019 | 3 | | 3 | | 33 | 33 | | |
| 12/26/2019 | 2 | | 2 | | 35 | 35 | | |
| 12/27/2019 | 2 | | 2 | | 37 | 37 | | |
| 12/29/2019 | 7 | 7 | 7 | | 44 | 44 | | |
| 12/31/2019 | 3 | | 3 | | 47 | 47 | | |
| 1/1/2020 | 10 | | 10 | | 57 | 57 | 1 | |
| 1/2/2020 | 6 | | 6 | | 63 | 63 | 1 | |
| 1/3/2020 | 12 | | 12 | | 75 | 75 | 3 | |
| 1/4/2020 | 12 | | 12 | | 87 | 87 | 5 | |
| 1/5/2020 | 15 | | 15 | | 102 | 102 | 8 | |
| 1/6/2020 | 17 | | 17 | | 119 | 119 | 10 | |
| 1/7/2020 | 30 | | 30 | | 149 | 149 | 12 | |
| 1/8/2020 | 43 | 8 | 51 | | 192 | 200 | 14 | |
| 1/9/2020 | 31 | 8 | 39 | | 223 | 239 | 17 | |
| 1/10/2020 | 32 | 4 | 36 | | 255 | 275 | 30 | |
| 1/11/2020 | 32 | | 32 | | 287 | 307 | 24 | |
| 1/12/2020 | 28 | 25 | 53 | | 315 | 360 | 29 | |
| 1/13/2020 | 21 | 30 | 51 | | 336 | 411 | 35 | |
| 1/14/2020 | 22 | 7 | 29 | | 358 | 440 | 42 | |
| 1/15/2020 | 22 | | 22 | | 380 | 462 | 50 | |
| 1/16/2020 | 12 | 41 | 53 | | 392 | 515 | 60 | |
| 1/17/2020 | 5 | | 5 | | 397 | 520 | 72 | |
| 1/18/2020 | 5 | | 5 | | 402 | 525 | 86 | |
| 1/19/2020 | 2 | | 2 | | 404 | 527 | 100 | |
| 1/20/2020 | 3 | 270 | 273 | | 407 | 800 | 123 | |
| 1/21/2020 | 2 | | 2 | 2255 | 409 | 802 | 143 | |
| 1/22/2020 | | | 854 | | | 1656 | 176 | |
| 1/23/2020 | | 340 | 340 | 3711 | | 1956 | | |
| 1/24/2020 | | | | | | | | |
| 1/25/2020 | | | | | | | | |
| 1/26/2020 | | | | | | | | 64960 |
| 1/27/2020 | | | | | | | | 75221 |
| 1/28/2020 | | | | | | | | 87484 |

# Appendix B.

# Timeline of COVID-19 Outbreak in Wuhan, before Lockdown on Jan 23, 2020

*Executive Summary*

- This timeline attempts to be comprehensive, with the main focus on "human-to-human" transmission
- By January 1, 2020, there had been convincing evidence of "human-to-human" transmission
- On January 20, 2020, Chinese communist party (CCP) government admitted "human-to-human" transmission including infection of medical staff
- Between January 1 and January 20, CCP government had been hiding the epidemic related information, suppressing the spread of epidemic related information and denying "human-to-human" transmission
- The delay of sharing with the world the true information related to the COVID-19 outbreak in Wuhan, leading to the largest public health and humanity crisis of this century

# Classification

**A**: Category A described how different layers of the Chinese Communist Party (CCP) government system use various strategies to hide the truth about Wuhan COVID-19 outbreak, including all measures and policies to suppress the right moves from the first line medical staff in Wuhan.

**B**: Category B described how the front line medical staff and other health care related professionals reacted and found out about the truth, and how they tried to manage the outbreak while blowing whistles.

Basically, the "human to human transmission" and "viral genome" are the key clues that we focus this timeline on.

# Timeline

September 18, 2019 (09/18/2019)

A: Chinese media reported a drill in Wuhan Customs. One case scenario was a case of patient with acute respiratory distress in Wuhan Airport, from a flight entering China. The patient was later diagnosed as "novel coronavirus infection".

http://k.sina.com.cn/article_2000016880_7735d5f002000krlq.html

11/17/2019

B: According to South China Moring Post, China's first confirmed Covid-19 case could be traced back to November 17, 2019. Government records suggest first person infected with new disease may have been a Hubei resident aged 55, but 'patient zero' has yet to be confirmed.

https://www.scmp.com/news/china/society/article/3074991/coronavirus-chinas-first-confirmed-covid-19-case-traced-back

12/1/2019

B: 1st recorded symptom onset of Wuhan virus patient was on 12/1/2019 from a paper Chinese medical expert published on <Lancet> with title of <Clinical features of patients infected with 2019 novel coronavirus in Wuhan, China>.

https://www.thelancet.com/journals/lancet/article/PIIS0140-6736(20)30183-5/fulltext

12/12/2019

B. Patient Huang, from the Seafood Wet market, presented to Wuhan Youfu Hospital which is about 200 meters from the Wet Market. Huang mentioned that seven to eight staff in his neighbor store got sick, with fever. Huang was concerned about himself. He did not follow doctor's suggestion to take a CT.

12/16/2019

B. Seventy years old male, presented to Youfu Hospital, with fever for several days. He took a chest x-ray which appeared no obvious abnormality.

（八点健闻）

12/16/2019

B. Patient A: Sixty-five years old, working as a delivery man in the Wet Market, presented to Wuhan Central Hospital, mentioning that he had persistent fever since Dec 13 (another saying Dec 15).

12/18/2019

B: Patient A with fever (same as above 12/16/19) presented to the ER of Wuhan Central Hospital. ER chief is Dr. Ai Fen. CT scan showed pulmonary infection with presentation of "multiple patchy blurry shadows in bilateral lungs". This patient was transferred to ICU and did not respond to multiple antibiotics. Bronchoalveolar lavage was performed on 12/24/2019 and samples sent to 3rd party diagnostics company Vision Medicals (广州微远基因科技有限公司) at Guangzhou for (NGS/mNGS) sequencing. Patient A was transferred to Wuhan Tongji Hospital on Dec 25.

12/26/2019

B: Patient Chen1 visited Wuhan Central Hospital. Alveolar lavage sample was sent to Shanghai Public Health Clinical Center affiliated to Fudan University.

12/26/2019

B: Prof. Zhang Yongzhen's team in Shanghai Public Health Clinical Center received one sample each from Wuhan Central Hospital and Wuhan CDC, respectively.

[[[ **From Wuhan Xinhua Hospital to Jinyintan Hospital: Series of events from 12/26-29, 2019**

(for the purpose of coherence, a few day's actions are put together):

12/26/2019

B: One patient from the Wet Market presented to Wuhan Xinhua Hospital. His symptoms included fever, cough, short of breath and "patchy ground glass" like lesions in the lungs.

12/27/2019

B: An elderly couple presented with Fever and dyspnea was admitted to Wuhan Xinhua Hospital ICU. ICU chief, Dr. Zhang Jixian reported these 3 similar cases to hospital leadership, hospital subsequently reported to local CDC in the morning. On 12/28/2019, the next day, son of the above elderly couple was called to Xinhua Hospital. The CT imaging of all three showed similar findings. **This family cluster demonstrated an obvious clue of "human to human transmission".**

12/27/2019

A: Local CDC came to Xinhua Hospital for epidemiological investigation in the afternoon and took history and samples from airway and blood.

12/28 and 12/29, 2019

B: There were another 3 patients with similar symptoms, all from the Wet Market.

12/29/2019

B: In the afternoon of 12/29, the vice president of Xinhua Hospital, Dr. Xia Wenguang, called for a clinical conference with 10 experts in different subspecialties for discussing these 7 patients. All experts concurred that these represented unusual situations. **Dr. Xia reported to the City and Provincial Health Commissions and CDCs. He asked for starting emergency protocol activation.**

A: In the same afternoon, Health commissions and CDCs from Wuhan City and Hubei Province organized an expert team to Xinhua Hospital.

B. 6/7 patients in Xinhua Hospital were transferred to Wuhan Jinyintan Hospital.

Patient A (the first one with sample submitted to NGS testing) from the Central also was transferred to Jinyintan Hospital.**]]]**

12/27/2019

B. Another patient (Chen2) presented to Wuhan Central Hospital. Alveolar lavage sample was sent to Beijing (CapitalBio MedLab, 北京博奥医学检验所有限公司) for NGS testing.

12/27/2019

B: The NGS sequencing results of patient A were orally reported back to Central Hospital by Vision Medicals by phone, it was claimed to be a coronavirus that is "Bat SARS like".

12/27/2019

B: Vice chief, Dr. Huang, of Jinyintan Hospital was told about patient A who was planned to be transferred to his hospital. **Dr. Huang obtained sequencing results of patient A from Vision Medicals and forwarded to Wuhan Institute of Virology (WIV). WIV replied with confirmation of high similarity to SARS-like coronavirus from bat**.

12/29/2019

B: ER in Wuhan Central Hospital reported 4 similar cases to Hospital Public Health Office, which reported to local county CDC. Among these 4 patients**, there was a pair of mother-son, another example of human-to-human transmission.**

12/29/2019

A: Local CDC and City Office arrived in Central hospital, surveyed and sampled the 7 patients.

12/30/2019

B: Wuhan Jinyintan hospital sent samples from patients to Wuhan CDC and Wuhan Institute of Virology (WIV). 2 out of 7 samples were identified to be SARS positive by Wuhan Institute of Virology on 12/30/2019.

12/30/2019

B: Wuhan Jinyintan Hospital received another 27 patients, with 13 as "severe". Regular floor wards were changed to "Quarantine wards".

12/30/2019

**B: Beijing CapitalBio MedLab (北京博奥医学检验所) reported sequencing results of patient chen 2 as "highly suspicious for SARS", to the Central Hospital.**

12/30/2019

B: Dr. Ai Fen, Chief in Central Hospital, reported sequencing result to Hospital Public Health Office and Hospital Infectious Disease Department. She also forwarded the report to ER doctor's Wechat.

12/30/2019

B: Dr.Li Wenliang from Wuhan central hospital shared information about SARS-like infection on Wechat and asked readers to warn family and friends. Dr. Liu and Dr. Xie also shared with their Wechat groups.

12/30/2019

Other whistle blower included Attending physician Dr. Xie Lingka in the Cancer Centerof Xiehe Hospital [（华中科技大学同济医学院附属）协和医院肿瘤中心主治医师谢琳卡].

12/30/2019

B: One physician in Xinhua Hospital reported symptoms. This was confirmed on Jan 11 and **she is the 2nd case of medical staff infection in Xinhua Hospital**.

12/30/2019

A: Wuhan City Health Commission issued 2 emergency policies, the 2nd one stated that "**no one or institution is allowed to disclose information about the pneumonia without authorization**."

12/30/2019

A: Wuhan City Health Commission broadcasted to major hospitals internally that "1, this pneumonia is not SARS, **2 no human to human transmission**, 3 does not cause death; 4 no medical staff infection". (see above right)

12/30/2019

B: Wuhan Central Hospital clinical departments shared the message/instruction from Wuhan City Health Commission not to "casually announce the current situation to public or outsider, otherwise they would be seriously punished".

12/30/2019

A: Leadership in Wuhan Central Hospital informed all division chiefs to inform their staff that they were not allowed to leak any virus related news and not allowed to wear facial masks.

12/31/2019

A: Chinese National Health Commission dispatched a panel of experts to Wuhan for investigation. Chinese CDC chief went to Wuhan on the earliest flight that day.

12/31/2019

B: Zhongnan Hospital in Wuhan initiated a special team for "Pneumonia with unknown etiology". The team leader asked all departments to prepare for infectious disease with human to human transmission potential but need to keep this confidential.

12/31/2019

B: Xiehe hospital in Wuhan established 24 beds in an isolation unit for cares of "Pneumonia with unknown etiology" only, and organized re-distribution of resources including staff.

12/31/2019

A: Dr. Li Wenliang was called to hospital administration office for his mistake on spreading "rumors" and was asked to provide written statement for his "mistake".

12/31/2019

A: Infectious control staff in PPE came to do cleaning in the Seafood Wet Market with sprayer.

12/31/2019

A: Wuhan media received messages that no reports were not allowed. They needed to follow official statements.

12/31/2019

A: Wuhan City Health Commission announced at 1:30 pm that recent pneumonia cases seen in some hospitals were related to the Wet Market; so far 27 cases were found, with 7 severe cases, rest cases were stable, and 2 cases improved and would be discharged soon; investigation suggested above were viral pneumonia; **so far no obvious findings of "human to human transmission" and no infection of medical staff.**

Note: For comparison (or see Table 3).

12/31/2019

A: China reported to WHO about a series of "Pneumonia with unknown etiology" in Wuhan.

12/31/2019

B: Taiwan inquired China and WHO for more information about the outbreak, asking for details about the "Atypical Pneumonia" "patients quarantined" with a strong suspicion of human to human transmission. Tianwan started border infectious screening on direct flights from Wuhan to Taiwan.

12/31/2019

Hong Kong media reported the presence of "new SARS" in Wuhan. Hong Kong government had press conference on the epidemic. They started almost daily press conference.

1/1/2020

B: A family clinic owner seeked medical care at Central hospital for pneumonia. ER chief Ai Fen contacted hospital for infection potential, her report was not answered. On the same day, she asked 200 ER staff to wear N95 masks, caps and wash hands. She also changed the surgical ward into an isolation unit with 20 beds.

1/1/2020

B. See Table 1. Summary: Evidence of "human to human transmission by 1/1/2020

1/1/2020

A: Huanan seafood market was ordered by the Wuhan City government to be closed.

1/1/2020

A: Chinese National Health Commission started a special team to manage the Wuhan outbreak with daily meetings. This was announced on 1/19/2020.

1/1/2020

B: Table 2. Marked increase of fever patients at new year of 2020.

1/1/2020

A: Table 3. Wuhan epidemiological data by 1/1/2020

1/1/2020

A：An anonymous person in Sequencing Company reported that he received a phone call from Wuhan Health Commission on 1/1/2020 that he had to stop testing all samples from Wuhan; destroy all existing samples; should not disclose information regarding samples to public; should not publish any research or data based on the samples from Wuhan. He was told to report to the Health Commission should the company receive any sample in the future.

1/1/2020

A: **Wuhan police department announced that Wuhan Health Commission has conveyed information about Wuhan pneumonia to the general public. However, some people spread false information online, 8 of them have been summoned and proper actions taken.**

1/1/2020

A: The first edition of "Diagnosis and treatment manual for viral pneumonia of unknown etiology" was proposed and distributed to Wuhan.

1/1/2020

A: WIV conducted the separation of virus from sample(s) provided by Jinyintan Hospital.

1/1/2020

B: **BGI group (Huada Gene) reported sequencing results of 3 cases of pneumonia which harboured a viral sequencing showing 80% similarity to SARS virus to Wuhan City Health Commission.**

1/2/2020

A：Wang Yanyi, Director of Wuhan Institute of Virology (WIV) emailed all staff to ensure they can not disclose any information about the new coronavirus.

1/2/2020

A：Dr. Ai Fen, ER chief in Wuhan Central Hospital, was seriously criticized by her superior at Central hospital for her "making rumors", she needed to verbally tell all her 200 staff not to disclose anything about this pneumonia. She was also told she can not use Wechat or text but only by face to face or phone call. Dr. Ai raised the issue of "human to human transmission" and she received no response.

1/2/2020

Wuhan Central Hospital leadership informed all staff, discussion of "pneumonia of unknown etiology" was not allowed except at the time of hands-off, any forms of documentation were not allowed.

1/2/2020

A: Wuhan Institute of Virology finished sequencing on 1/2/2020 of COVID-19, which was characterized as a "novel" coronavirus, the results were presented to GISAID on 1/11/2020 and publicized on 2/4/2020.

1/2/2020

A: **CCTV13 aired the news that eight people in Wuhan were punished by "making and spreading rumors".**

1/3/2020:

A: Wuhan police had a meeting with Dr. Li Wenliang regarding his mistake of "spreading rumors". (see below Admonition with his finger stamp):

1/3/2020

A: Central hospital administration (president, party secretary and party discipline commission secretary) criticized division chiefs for wearing masks. Anonymous physician mentioned that Dr.Li Wenliang was about to be fired by the hospital after he was reprimanded by the hospital admin. Hospital admin also issued commands verbally such as "can not tell", "can not wear masks". Hospital admin agreed for the ER, respiratory unit and ICU staff to wear N95 masks but asking other departments not to wear any masks, which lead to massive medical staff infection due to exposure to virus without protection in central hospital: Up to Feb-11, more than 230 medical staff were diagnosed with COVID-19 infection and up to March-09, 4 physicians died, including Dr. Li Wenliang.

1/3/2020

A: Wuhan Health Commission : So far 44 cases with 11 severe cases. **No evidence of "human to human transmission", no infection of medical staff.**

1/3/2020

B: Wuhan reported pneumonia through an "online direct reporting system" for the first time to Chinese CDC. Chinese CDC former vice chief Yang Gongchuan stated that she asked the national CDC and confirmed that there was no "direct report" from Wuhan in 2019. Chinese CDC mentioned receiving a direct report from Wuhan from 1/3/2020 -1/10/2020, but reporting stopped suddenly till 1/24/2020.

1/3/2020

B: Chinese National Health Commission published 1st protocol for pneumonia with unknown viral etiology for Wuhan Health Commission to distribute.

1/3/2020

A: Wuhan Health Commission made 2 versions of the protocol with 2 covers. Protocol with Green cover has 9 files in it while the one with white cover has 10 files. Wuhan Health Commission distributed protocol with white cover with very strict diagnostic criterias. The green protocol was given to Chinese National Health Commission but the white protocol was used for medical practice in Wuhan. The fundamental difference is that "contact with Wuhan seafood wet market or contact history with suspected patients" is a must for diagnosis in the white protocol but optional in the green protocol. Following the white protocol will lead to massive underdiagnosis, under testing and under reporting.

1/3/2020

A: Chinese CDC Institute of Virology finished sequencing the whole genomic sequence for the novel coronavirus. This sequencing information were kept as secret.

1/3/2020

A: **Chinese National Health Commission** published "Document on 3" (or Prohibition on 3rd) on 1/3/2020 to **prohibit any organization from publishing information about pathogen detection and test results, asking all organizations to destroy virus samples or transfer to designated institutions. Any organization of individuals that did not follow this policy will be legally punished.**

1/3/2020

A: Chinese government started to update the US government about outbreak information, but not to the Chinese public.

1/4/2020

Hong Kong government announced the eresponse plan for the novel infectious disease of public health significance.

https://www.chp.gov.hk/files/pdf/govt_preparedness_and_response_plan_for_novel_infectious_disease_of_public_health_significance_chi.pdf

1/5/2020

B: Professor Zhang Yongzhen team from Shanghai Public Health Clinical Center detected a novel SARS like coronavirus and obtained the whole genomic sequence from their samples obtained on 1/3/2020. **Zhang notified Shanghai Health Commission and National Health Commission to warn them that the new virus is in the same origin with SARS and spread from the respiratory tract and recommended disease control measures in public.**

1/6/2020

A: **One physician in the Pulmonary Medicine department has abnormal findings in his lungs in Wuhan Xinhua hospital**. The hospital held an internal meeting which stressed that it was not allowed to leak this information out especially "can not tell the media".

1/6/2020

A: Wuhan "Two sessions" started. No public media updated the public about the outbreak.

1/6/2020

A: **Chinese National CDC started 2nd degree emergency response internally.**

1/7/2020

**Chinese President Xi asked that the anti-epidemic measures should not affect normal life.**

http://www.qstheory.cn/dukan/qs/2020-02/15/c_1125572832.htm

1/9/2020

A: Insider reported to Caixin Media that Zhongnan Hospital public health department submitted 2 pneumonia cases on 1/9/2020 with **direct reporting system to Chinese National CDC and had system confirmation of successful submission.** However, they found that the submission on 1/9/2020 **was deleted**. Wuhan CDC staff confirmed with Caixin Media that the Health Commission and CDC on district, city and province levels can see submission of cases from local hospitals in the system but district and city level do not have the authorization to delete them.

1/9/2020

In Hong Kong, experts assumed that the epidemic could have "human to human transmission".

1/10/2020:

A: Wuhan "Two sessions" ended, no information of the epidemic was discussed at the conference.

A: Wuhan Health Commission announced its 1st epidemic report. It mentioned that **no new cases were identified since 1/3/2020. No evidence for human to human transmission and no medical staff was infected.**

A: Dr. Wang Guangfa, expert team member for the National Health Commission, said that the outbreak is controllable when interviewed.

1/10/2020

B: Radiologist, Dr. Li at Wuhan Xinhua hospital identified 3 cases from CT reports with ground-glass opacity. Similar cases increase daily with an exponential manner. There were 30 cases on 1/10/2020. Dr. Li believed that situation was severe since he had never seen any virus that grows this fast and multiplied every few days. Dr.Li stopped believing official reports and contacted his peers from other radiology departments, the situations are not optimistic across hospitals.

1/11/2020

B Prof. Zhang Yongzhen's team from the lab of Shanghai Public Health Clinical Center **published the genetic sequence of COVID-19 to GISAID and used GISAID、virological.org and GenBank to share with the world** at 9am on 1/11/2020. Chinese National Health Commission announced that China will share the virus genetic sequence with WHO that evening.

1/11/2020

A：Chinese National Health Commission published other 5 genetic sequences from different patients, those 5 sequences are from Chinese CDC, Chinese medical science institute and Wuhan Institute of Virology. (Wuhan Institute of Virology confirmed the sequence on 1/2/2020 but upload on 1/11/2020 to GISAID.)

1/11/2020

Publication on NEJM indicated that there were 7 medical staff infected during Jan1-11, 2020.

https://www.nejm.org/doi/full/10.1056/NEJMoa2001316

1/12/2020

A: Shanghai Health Commission ordered the lab (Prof. Zhang Yongzhen) in Shanghai Public Health Clinical Center, who submitted viral genome sequence to GISAIDon 1/11/2020, **to close down and subject to disciplinary actions**, as the consequence of violating the "3rd Document" (or Prohibition on 3rd) issued on 1/3/2020.

1/12/2020

A: Wuhan CCP secretary, Ma Guoqiang stated on CCTV interview on 1/31/2020 that Wuhan had an increase in pneumonia cases on 1/12 and 1/13. Thus Wuhan started to measure passenger temperatures at airports and high-speed train stations.

1/12/2020

A: The reporting for pneumonia cannot be submitted without approval from Hubei Health Commission, thus reporting was artificially stopped from local hospitals for 5 days.

1/14/202

A: Director of Chinese National Health Commission had video conference with provincial directors.

1/15/2020

A: Chinese National Health Commission held video conferences on prevention and control of the epidemic. Exposed internal confidential documents (see below link) revealed that the meeting **conveyed five messages**: 1). Alarm of the epidemics. 2). Policy is "keep tension internally and keep a relaxed appearance to public". 3). Prepare for the epidemic. 4). Ensure the strict procedure for diagnosis of first case to be followed, i.e., the first case can only be made by the national level and announced by the province level. 5). Reduce the fatality rate since death would lead to public panic.

https://www.epochtimes.com/gb/20/6/2/n12156428.htm

1/17/2020

B: Dr. Yuen Kwok-yung, a Hong Kong microbiologist, visited Wuhan with a panel of experts. Yuen felt the places he visited were all prepared for the Q&A sessions. Yuen later **asked Guangdong CDC to prepare for human to human transmissions.** Chief of Zhongnan hospital ICU protested to the expert panel that "the diagnostic criteria were too strict and will omit true patients."

1/18/2020

A: Wuhan Health Commission reported 4 new cases. However, it reported no new cases from 1/12-1/17. Multiple hospitals had the same experience of not being able to report during this period.

1/18/2020

B: Wuhan community organized group dinner event - "Ten-thousand families dinner" in which forty-thousand families participated.

1/19/2020

A: Wuhan CDC and government stated to the media that Wuhan novel coronavirus had low infectivity and can be controlled and prevented.

1/19/2020

B: Dr. Li observed in the reading room that "in the beginning there were 2-3 cases daily, 4-5 cases on day 2, 7-8 cases on day 3, there was no obvious growth in the first 3 days. Then it entered an exponential growth phase, about 30 cases on 1/10 then doubled every 3-4 days, 86 cases on 1/18 then above 100 everyday afterwards.

1/20/2020

B: Jinyintan, Hankou and Wuhan Pulmonary hospital became 3 designated hospitals for pneumonia with a total of 800 beds created. Other hospitals provided 1,200 beds, however the 2,000 beds were soon all occupied.

A: **Dr. Zhong Nanshan confirmed "human – to – human" transmission and infection of health care professionals** at the press conference.

1/21/2020

**Wuhan Health Commission announced the infection of 15 medical staff.**

1/22/2020

A: **Chinese National Health Commission reported that there was human to human transmission and infection of medical staff**, source of virus and how virus spread was not identified.

1/22/2020

A: Gao Fu, director of Chinese CDC, commented in press conference that the origin of the epidemic was wild animals.

1/22/2020

B: Fever clinic visits for Red Cross hospital reached 1,700 on 1/22/2020 given the majority of hospitals closed their fever clinics on that day . Visits reached 2,400 the day after at that hospital.

1/23/2020

A: Lockdown of Wuhan was announced. Five million Wuhan residents left Wuhan before lockdown.

# Appendix B.

Table 1. Summary: Evidence of "human to human transmission by 1/1/2020

| Hospital | Case | Dates | Action | Reference /media reports |
|---|---|---|---|---|
| Xinhua Hospital | The case of family of 3 | Dec 27 and 28, 2019, in addition to the 4 cases from the Wet Market (total 7) | Hospital reported to District CDC 12/27/19; | [1] |
| Central Hospital | Sequencing result of "Bat SARS like" novel coronavirus reported by Guang Zhou (Vision Medical) from patient A (Dec-18-2019) | Admitted on Dec 18, 2019; Sampled on Dec 24, 2019; Sequencing result reported on December 30, 2019 | | [2] |
| Central Hospital | The case of mother-son pair | December 29, 2019 | | [2] |
| Central Hospital | Beijing (CapitalBio MedLab) reported sequencing results of patient Chen 2 as "highly suspicious for SARS". | Admitted on December 27, 2019; Reported on December 30, 2019 | | [3], [4] |
| Xinhua Hospital | One physician in Xinhua Hospital reported symptoms. | December 30, 2019 | Xinhua Hospital | |

| | This was confirmed on Jan 11, 2020 and she is the 2nd case of medical staff infection in Xinhua Hospital. | | | |
| --- | --- | --- | --- | --- |
| Central Hospital | A patient who was the owner of a family clinic at the Wet Market | Jan 1, 2020 | | [2] |
| Summary of sequencing results | By December 30, 2019, samples were taken from no less than 9 patients of "pneumonia of unknown etiology" in Wuhan, including 6 from the Xinhua Hospital. Sequencing results suggested that the pathogen is a SARS-like coronavirus. These results were reported back to hospitals and reported to Health Commission and CDC. | | | [4] |

Note:

# Appendix B

Table 2. Wuhan epidemiological data by 1/1/2020 according to different resources

| Presenter | Nov-17-19 | Dec-1-19 | Dec-8-19 | Dec-31-19 | Jan-1-20 | Methods | Note |
|---|---|---|---|---|---|---|---|
| Wuhan Health Commission | | | Frist case as announced on Jan-11-20. | 27 | | Official announcement | |
| Huang et al, Lancet | | 1 | | 41 | | | [1] |
| Epidemiology Working Group for NCIP Epidemic Response, Chinese Center for | | | | 104 | | Extracted from China's Infectious Disease Information System. | [2] |
| South China Morning Post | First case of COVID-19 confirmed case can be traced back to this day. | | 27 by Dec-15, 60 by Dec-20, 180 by Dec-27, 2019. | 266 | 381 | Internal data from Chinese government | [3] |
| Li Q et al | | | | 47 | | "pneumonia of unknown etiology" surveillance mechanism | [4] |

Note:

[1]. https://www.thelancet.com/journals/lancet/article/PIIS0140-6736(20)30183-5/fulltext

[2]. https://pubmed.ncbi.nlm.nih.gov/32064853/

[3]. https://www.scmp.com/news/china/society/article/3074991/coronavirus-chinas-first-confirmed-covid-19-case-traced-back

[4]. https://www.nejm.org/doi/full/10.1056/NEJMoa2001316

Table 3. Early sequencing of viral genomes by different labs

| Sequencing lab | Patient | Hospital | Note |
| --- | --- | --- | --- |
| Vision Medical, (广州微远基因科技有限公司) Guang Zhou; Collaborator is Chinese Medical Science Academy Institute of Virology [1] | Patient A | Wuhan Central Hospital. Admitted on Dec 18, 2019; Sampled on Dec 24, 2019; Sequencing result reported on December 30, 2019 | "Bat SARS like" novel coronavirus. Collaborator is Chinese Medical Science Academy Institute of Virology which submitted the sequence to GISAID in the evening on 1/11/20. |
| Beijing (CapitalBio MedLab) [2] | Patient Chen 2 | Wuhan Central Hospital. Admitted on December 27, 2019; Reported on December 30, 2019 | reported sequencing results of as "highly suspicious for SARS". |
| Shanghai Public Health Clinical Center affiliated to Fudan University [3] | Patient Chen 1 | Wuhan Central Hospital. Sequencing finished on 1/5/20. | First lab submitted COVID-19 sequence to GISAID at 9 am on 1/11/20. Lab was suspended on 1/12/20. |
| Wuhan Institute of Virology, Chinese Science Academy [4] |  | Sampled on December 30, 2019, Jinyintan Hospital | Finished viral genome sequencing on 1/2/20; submitted to GISAID on 1/11/20 and made public on 2/4/20. |
| BGI groups [5] | 3 positive samples from 30 samples from Wuhan in December, 2019 | Multiple hospitals 3 positive samples received on 12/16, 12/29 and 12/30 | 3 positive cases showing high similarity to SARS (> 80%); SARS testing kit negative; reported |

|  |  |  | to Wuhan Health Commission on 1/1/2020; whole genome sequencing on 1/3/2020. |
|---|---|---|---|
| Chinese CDC [6] | Finished sequencing on 1/3/20. | Received 4 samples from Wuhan on 1/2/2020. | Viral genome sequence submitted to GISAID on 1/12/20. |